\documentclass[11pt,a4paper]{article}
\usepackage{graphicx}
 \usepackage{epstopdf}
\usepackage{amsmath}
\usepackage{amssymb}
\usepackage{mathrsfs}
\begin{document}

\title{ Accurate prediction of the glass transition in classical fluids:   a pragmatic modification of the mode coupling theory\\
}
\author{S. Amokrane $^{1,*}$, F. Tchangnwa Nya $^{1,2,3}$, and J.M. Ndjaka $^{3}$ \\
1: Physique des Liquides et Milieux Complexes,
Facult\'{e} des Sciences et Technologie, \\
 Universit\'{e} Paris-Est (Cr\'{e}teil), 
61 Av. du G\'{e}n\'{e}ral de Gaulle, 94010 Cr\'{e}teil Cedex, France \\
2: Universit\'{e} de Maroua, Institut Sup\'{e}rieur du Sahel, Dpt. du Traitement des Mat\'{e}riaux,\\
 Architecture et Habitat, BP 46, Republic of Cameroon \\
3: D\'{e}partement de Physique, Facult\'{e} des Sciences, Universit\'{e} de Yaound\'{e} I. B.P. 812, \\
Yaound\'{e}, Republic of Cameroon\\} 

\date{17/07/2012}
\maketitle
\begin{abstract}
Many qualitative observations on  the glass transition  in classical fluids are well described by the mode coupling theory (MCT) but  the extent of  the non-ergodicity domain is often over-estimated by this theory. Making it more quantitative while  preserving its microscopic nature remains thus a current challenge. We propose here a simple heuristic modification that achieves this for the long-time limit quantities by  reducing the  excess of static correlations  that are presumably responsible for its inaccuracy. The location  of the ideal  glass transition predicted from this modified MCT compares very well with simulation for a wide range of interaction potentials  in pure fluids and  in mixtures.  
\end{abstract}

Being based on a time evolution equation deduced from a true hamiltonian, MCT is one of the few microscopic theories of the glass transition \cite{Gotze}. Developed initially for atomic fluids,  its has more recently regained interest  for describing the ideal glass transition in soft condensed matter \cite{Szamel,Dawson}. While it reproduces  important qualitative observations in both physical situations,  it suffers from quantitative limitations due to its approximate treatment of the time evolution equation. Examples are the critical packing fraction $\eta_g^{MCT} =0.525$ of the hard-sphere (HS) glass instead of $\eta_g^{ex} \approx0.58$ from experiments on HS colloids \cite{Pusey}, the critical temperature $T_g^{MCT} $ for the Lennard-Jones (LJ) fluid being in error by a factor two \cite{Bengtzelius} and the whole  $T_g^{MCT}(\rho)$ curve  for the square-well (SW) fluid \cite {Sciortino} clearly misplaced in the temperature-density plane. Some ad-hoc recipes have been proposed to correct this  in specific situations such as the LJ mixture \cite {Nauroth},  but they are not completely satisfactory.  Comparison with experiment or simulation is usually done by rescaling the state variables (see for example \cite {Sciortino}), but  at the expense of the  microscopic nature of the theory \cite{Berthier,Biroli}.  To go beyond this, one should start from the basic approximations of  MCT.  Besides the projection of the dynamical variables on  specific subsets, an important one is the factorization of four point contributions as products of pair terms\cite{Gotze,Szamel}.  The static structure appears then in MCT at the level of  the structure factors $S_{\alpha\beta}(q)$ and the triplet direct correlation functions (dcfs) $c_{\alpha\beta}^{(3)}(q)$ (in Fourier space).  The latter can  usually be neglected (convolution approximation), as shown by previous estimates \cite{Barrat-C3} and confirmed by more recent ones \cite{Abd-C3} . As the liquid-glass transition is  not very sensitive to the quality of the static structure, at least for hard-spheres \cite{Fridolin}, it is tempting to view this overestimation  by MCT of the non-ergodicity domain as reflecting too strong pair correlations, and attribute this to the factorization approximation. They should accordingly be reduced (the possibility to predict the dynamics only from the static structure has recently been  criticized \cite{Berthier}. However, the factored  four point terms involving two time-dependent densities in $\bold{q} $ space,  it is understandable that this  approximation is  more critical for  the time dependence since different time scales might be improperly mixed).  In this letter, we propose and test a pragmatic way of implementing this idea in the long-time limit: to reduce  correlations, one should compute the static structure at an effective density  that is lower (or a temperature higher) than the actual one (see  \cite{Nauroth} for a similar attempt). To see how this can be done in practice,  recall that  the densities $\rho_\alpha=\frac{N_\alpha}{V}$ enter the MCT time evolution of  the matrix $\bold{S}(q;t) $ with elements $S_{\alpha\beta}(q;t)$ 
\begin{equation}
 	\frac{\partial}{\partial t}\bold{S}(q;t)+q^2\bold{H}(q)\bold{S}^{-1}(q)\bold{S}(q;t)+\int^t_0 dt'\bold{M}(q;t-t')\bold{H}^{-1}(q)\frac{\partial}{\partial t'}\bold{S}(q;t')=0
\end{equation}
 through  the irreducible collective memory function whose matrix elements are \cite{Nagele1,Note}:
\begin{equation}
 	M_{\mu \nu}(q,t)=\frac{D^0_\mu D^0_\nu}{16\pi^3(\rho_\mu\rho_\nu)^{(1/2)}}\sum_{\gamma \gamma' \delta \delta'} \int d{\bf k}V_{\mu;\gamma \delta}({\bf q,k})V_{\nu;\gamma' \delta'}({\bf q,k})S_{\gamma \gamma' }(\|{\bf q}-{\bf k}\|;t)S_{\delta \delta'}(k;t)
\end{equation}
 		In  the vertex amplitude,
		 \begin{equation}
 	V_{\mu;\gamma \delta}({\bf q},{\bf k})=\frac{1}{q}[{\bf q}.{\bf k}\delta_{\mu \delta} C_{\mu \gamma}(k)+{\bf q}.({\bf q-k})\delta_{\mu \gamma} C_{\mu \delta}(\|{\bf q}-{\bf k}\|)+q^2x_\mu^{1/2} C^{(3)}_{\mu \gamma \delta}({\bf k},{\bf q}-{\bf k}) ]
 \end{equation} 
 the static dcfs are computed from standard methods of liquid state theory with the actual densities ${\rho_\alpha}$ as input. The modification consists then in using  effective densities ${\rho_\alpha^{eff}}$ (specified below) in the vertex while retaining in $M_{\mu \nu}$ the explicit dependence on density in the factor before the integral.  As a result, the MCT equation is solved for the actual densities but with a static structure computed for ${\rho_\alpha^{eff}}$. Similarly, an effective temperature $T^{eff}$ is used when temperature is a relevant variable. It is stressed that this is not a mere rescaling of the variables, a posteriori. Rather, this amounts to solving a modified  MCT equation with an "effective " pair structure. To distinguish the results relative to the original MCT from those obtained from the modified MCT, we shall label the latter by a tilde (the label ex will refer to  experiment,  mostly  computer ones here). 

To calibrate the correction, we start from the hard sphere potential for which there is a purely repulsive glass (caging mechanism). From the difference $\triangle\eta=0.055$ between the experimental critical glass packing fraction $\eta_g^{ex}= 0.58$ and the MCT one $\eta_g^{MCT}= 0.525$ with accurate static input \cite{Bengtzelius,Abd-C3}, we solve the MCT equation for the non-ergodicity parameter $f(q)$ - at the packing fraction $\eta$ - by using the structure factor $S(q;\eta^{eff})$  for an effective packing fraction $\eta^{eff}=\eta-0.055$.  The solution of this modified integral equation is the standard method of direct iterations \cite{Gotze}. The first non trivial result is that these iterations converge to a new critical value $\tilde{\eta}_g^{MCT} =0.584$, virtually the exact one (ie $\eta_g^{MCT}+\triangle\eta) $. Equipped with this calibration, we consider a binary hard-sphere mixture (size ratio $\delta=D_1/D_2$, packing fractions $\eta_1$ and $\eta_2$). The effective packing fractions are then  $\eta_i^{eff}=\eta_i-\triangle\eta(1-\eta_j/\eta_g^{ex})$; $i\neq j=1,2$. The slight dependence on the packing fraction of the other species is introduced  so as to recover the one-component correction  and prevent negative value. As with the one component case, we found that the modified MCT equations  converge nearly  to the corrected density, up to a size ratio $\delta=0.1$. Quantitatively, the correction was tested on  two HS mixtures considered in ref. \cite{Foffi-Voigt}. For  $\delta=0.6$ and $\hat{x}=\eta_1/\eta=0.2$, the total critical packing fractions are $\eta_g^{MCT}= 0.528$, $\tilde{\eta}_g^{MCT}= 0.593$, while  $\eta_g^{ex}= 0.606$ from simulation. For $\delta=0.83$ and $\hat{x}=0.37$, $\eta_g^{MCT}= 0.524$, $\tilde{\eta}_g^{MCT}= 0.589$ and $\eta_g^{ex}= 0.586$. For both mixtures,  the values predicted by the modified MCT are thus very close to  simulation. This definitive  improvement is confirmed by figure 1, which shows, as an example, the non ergodicity parameter $f_{11}(q)$ for $\delta=0.6$. This excellent agreement with simulation - at the new critical density-  is consistent with the relation  $\eta_g^{eff}\approx\eta_g^{ex}-\triangle\eta$.  It reflects  the fact that the original  MCT predicts the correct critical non-ergodicity parameter but a slightly inaccurate  critical density, as found in \cite{Fridolin}. 

To go beyond the HS model,  we first took the soft sphere potential $\phi=\epsilon(\frac{\sigma}{r})^{12}$ for which  density and  temperature are combined in the coupling constant $\Gamma=\rho\sigma^3(\epsilon/k_BT)^{1/4}$.  The one-component fluid of soft spheres has a critical coupling constant $\Gamma_g^{ex}=1.5$ (see \cite{Bernu}). Using the Rogers-Young closure for the static structure as in ref. \cite{Barrat-Mel}, one gets $\Gamma_g^{MCT}=1.33$. Using as the only information the one gained from the HS potential, we compute the  static structure in the modified MCT with an effective coupling constant $\tilde{\Gamma}=\Gamma-1.33(0.58/0.525-1)$. We then find  a new critical coupling constant $\tilde{\Gamma}_g^{MCT}=1.51$ again in very good agreement with  simulation. We next took the mixture considered by Barrat and Latz. One finds $\Gamma_g^{MCT}=1.32$  significantly lower than the simulation value $\Gamma_g^{MCT}=1.46$ ($\sigma^3 $ is replaced by $\sigma_{eff}^3=x_1^2\sigma_{11}^3+ 2.x_1x_2\sigma_{12}^3+x_2^2\sigma_{22}^3$ in the definition of $\Gamma $). With the modified MCT, the improvement is clear:   $\tilde{\Gamma}_g^{MCT}=1.51.$  (using .575 instead of .58, we  get $\tilde{\Gamma}_g^{MCT}=1.48$ virtually the exact result).

To consider a model with attractive contributions, we took the LJ fluid for which simulation data have been collected in \cite{Bengtzelius} and more recently in \cite{Dileonardo}.   Temperature plays now a role (the reduced temperature $T^*=k_BT/\epsilon$ is used): a large error in the  critical temperature $T_g^{MCT}(\rho)$ can arise  due to its very steep variation with $\rho$ : a $10\%$ variation of the density changes $T_g$ by a factor 2  \cite{Bengtzelius,Nauroth}. In this situation, slight details in the static structure might become relevant. The previous corrections for hard and soft spheres can be adapted by first noticing that the critical temperature $T_g(\rho)$ is well estimated from the critical packing fraction of hard spheres $\eta_g=\frac{\pi}{6}\rho d^3_{HS}(T,\rho)$ where $d_{HS}(T,\rho)$ is a suitably defined hard-sphere diameter  (see eg. \cite{Bengtzelius} and  \cite{Robles}).  In this spirit, we found (figure 2) that an even simpler formula $\frac{\pi}{6}\rho d^3_{HS}(T)=0.58$ fits very well the data of ref. \cite{Bengtzelius} when the diameter is taken as the distance at which the reduced potential has a value $\ln(a)$ \cite{Stell}: $d_{HS}(T)=\sigma[\frac{2}{1+\sqrt{aT^*}}]^{1/6}$, with $a=0.8$. $T_g^{MCT}(\rho)$ is also well fitted by a similar law with $a=0.5$ for the effective diameter $d_{eff}(T)$. Considering this as a seed, we computed the structure with an effective packing fraction $\eta^{eff}=\eta-\triangle \eta$ with  $\triangle \eta= 0.58(1/d^3_{HS}(T)-1/d^3_{eff}(T))$.

 The results is shown in figure 2. With the proposed modification,  $\tilde{T}_g^{MCT}(\rho)$ falls nearly exactly on the fitted law. As a check of the sensitivity to the choice of  $\triangle\eta $, the equation was solved by fixing it  to the value for $T_g^{MCT}(\rho)$ (ie the critical temperature predicted from the original MCT, for each density $\rho$). The results shown by  triangles are close to the simulation of \cite{Dileonardo} which differ slightly from  those of refs \cite{Bengtzelius} (this difference is clearly visible on the enlarged scale used in figure 2). This  might reflect different characterization of the glass (see e.g. \cite{Voigtmann_prl}).  The closeness  of the triangles with the simulations of \cite{Dileonardo}  is probably fortuitous, and there is a priori no reason to keep  $\triangle\eta$ fixed.  With the definition of $\eta^{eff}$ indicated above ( hence with a temperature dependent $\triangle\eta$ ),  we considered a standard LJ mixture at  the state point studied in  \cite{Nauroth}: one finds $T_g^{MCT}=0.922$ more than twice the simulation value $T_g^{ex}=0.435$. With the modified MCT, we get $\tilde{T}_g^{MCT}=0.56$, again  definitively better than the original one (to speed the calculations, the static structure was computed from the closure of ref. \cite{Duh}). 
 
 Encouraged by  these results, we finally considered the square well potential with short interaction range, a prototype for colloidal glasses for which an additional mechanism of dynamical arrest - the attractive glass - has been evidenced \cite{Dawson}. In this case,  the MCT glass  lines can be superimposed to simulation only with the help of a double linear transformation of both  temperature and density \cite{Sciortino}. In figure 3, we show the result from the modified MCT equation  using a temperature independent correction $\triangle\eta$,  a temperature dependent one $\triangle\eta(T) $ with a linear dependence on $1/T^*$ adjusted from a single simulation point besides the HS value, and lastly both  $\triangle\eta(T) $ and an effective temperature $T^*_{eff}=T^*+0.18$ ( to prevent cristallization, the simulation are actually for a slightly asymmetric mixture while we  solved  the MCT for the one-component fluid in order to compare with ref. \cite{Dawson}).  Even in this more complex situation in which the mechanism for the arrest involves the formation of long-lived bonds,  this way of reducing correlations leads to a clear improvement  over  the original MCT predictions.

In conclusion, independently of the correctness of the view that  attributes the quantitative insufficiency of MCT to the factorization approximation, there is  little doubt that the method proposed here for tempering correlations definitely improves the accuracy of the present form of MCT, insofar as the long time results are concerned. Besides the operational value of this modification for accurately predicting the ideal glass transition for quite disparate potential, it is expected that these results that  emphasize again the importance of many-body correlations and the presence of general mechanisms common to a wide class of models  will stimulate studies from first principles. In the framework of microscopic theories, the extension of this idea to the time dependent quantities, by taking account different time scales, is for example conceivable. This should go in parallel with the development of more global approaches. Finally, this significant improvement of the MCT predictions  that are relative to an idealized description of the glass transition  should be useful in a step by step progress towards a better understanding of the complex phenomena that occur  in real glass forming liquids. \\

* \textbf{Author for correspondence}:
amokrane@univ-paris12.fr\\
The authors are grateful to A. Ayadim and Ph. Germain  for useful discussions

\newpage
\setcounter{equation}{0} 

\section*{Figures}
\begin{figure}[!h]
\centering\includegraphics[width=7cm]{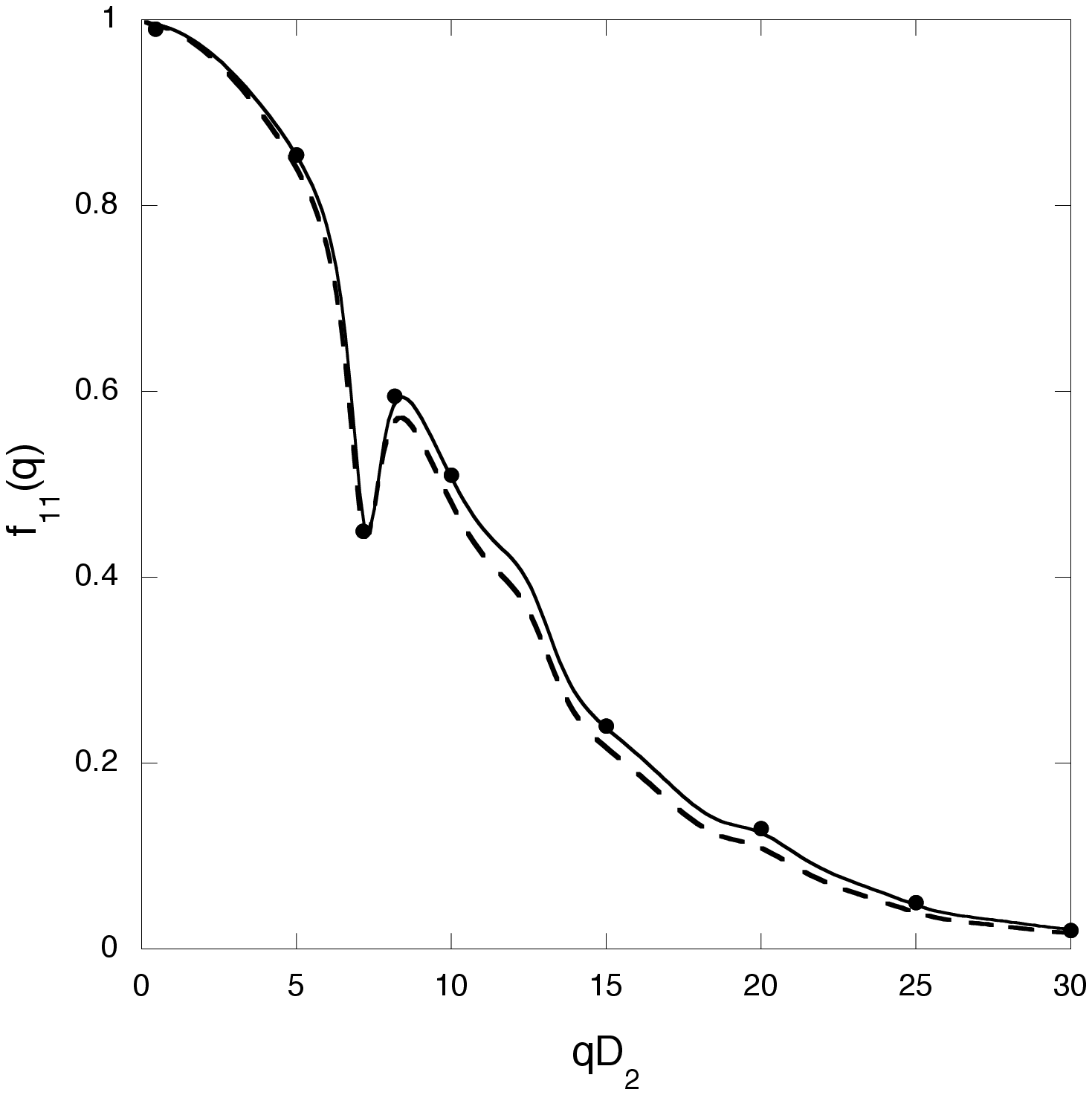}
\caption{Non-ergodicity parameter $f_{11}(q)$ in a binary hard-sphere mixture with $D_1/D_2=0.6$ and $\hat{x}=0.2$ at the glass transition,  Solid line: modified MCT ($\tilde{\eta}_g^{MCT}=0.593$);  dashes: original MCT \cite{Fridolin} ($\eta_g^{MCT}=0.5275$); Symbols: simulation \cite{Foffi-Voigt} ($\eta_g^{ex}=0.606$).}
\end{figure}

\begin{figure}[!h]
\centering\includegraphics[width=21cm,height=21cm]{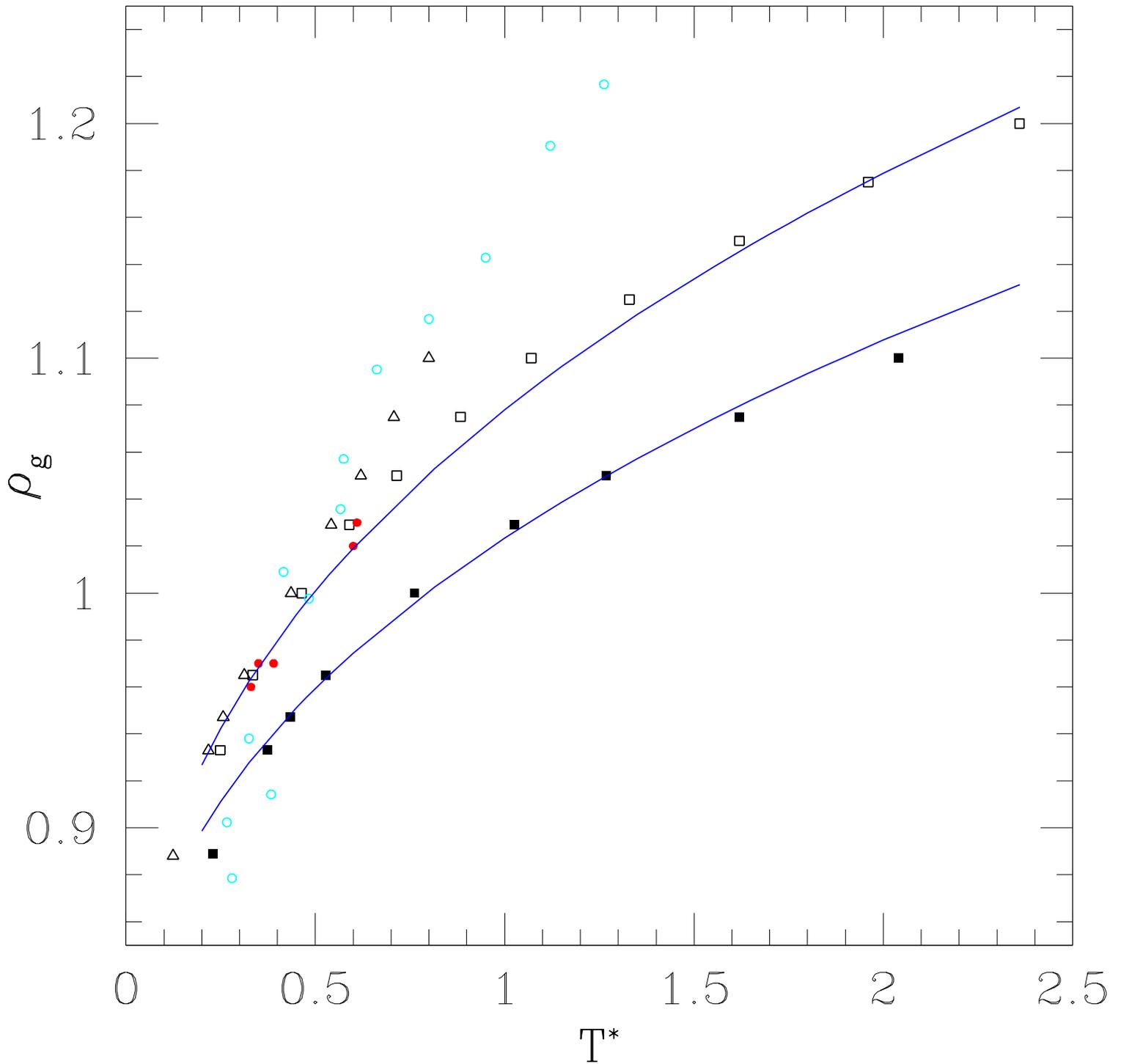}

\caption{Glass transition line for the Lennard-Jones fluid.} Filled squares and curve: original MCT and fitted curve with $a=0.5$ in the effective diameter; Empty squares: modified MCT and curve adjusted to the simulation of \cite{Bengtzelius} (filled circles) with $a=0.8$; triangles: Modified MCT with  fixed $\triangle\eta$; empty circle: simulations of \cite{Dileonardo}
\end{figure}

\begin{figure}[!h]
\centering{\includegraphics[width=7cm]{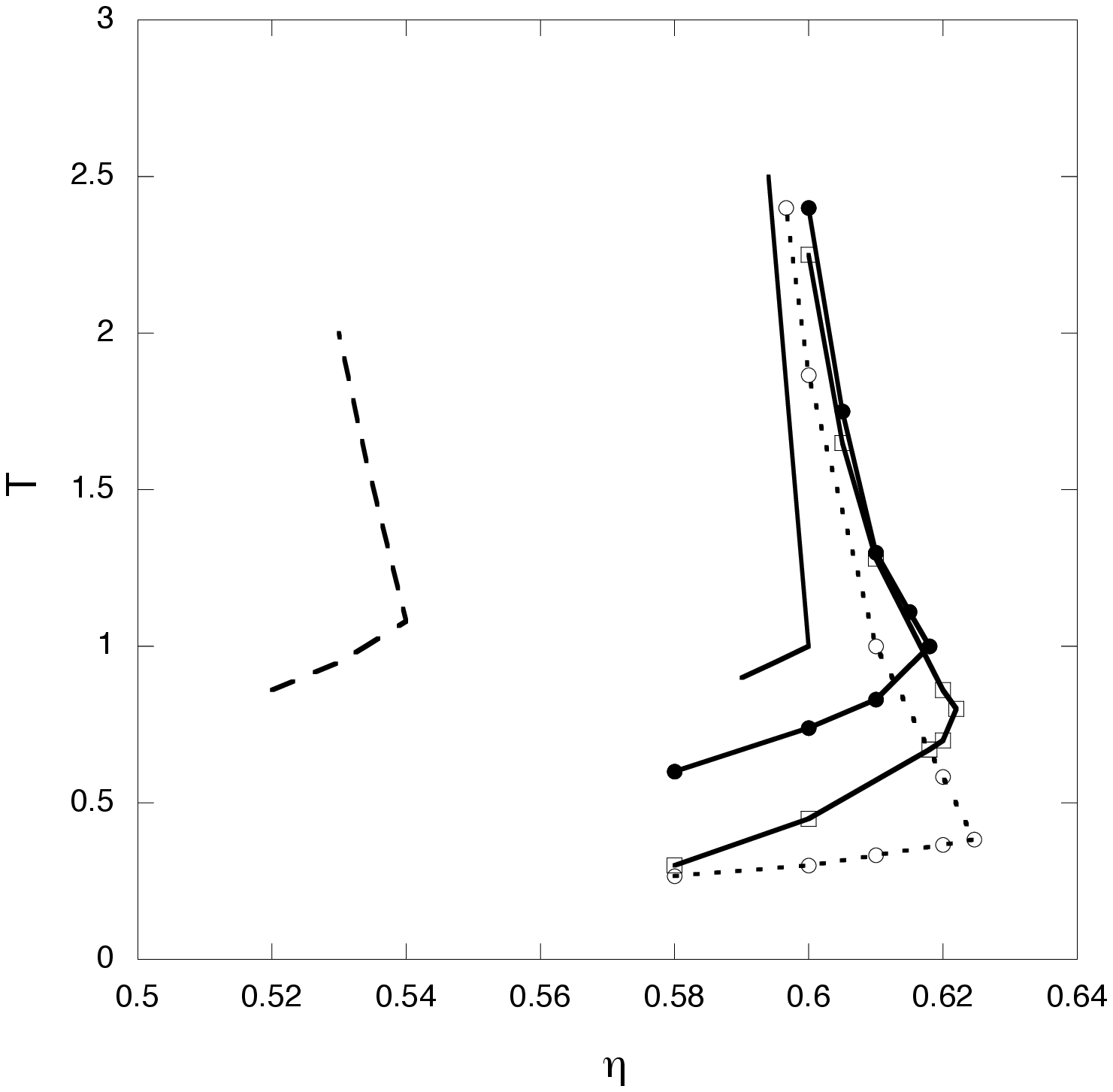}}
\caption{predicted glass transition lines for the square well fluid.} Dashes: original MCT; solid line without symbols: modified MCT with $\triangle\eta=0.055$; filled circles: $\triangle\eta(T) $; squares : $\triangle\eta(T) $ and $T^{eff}$. The theoretical results are for the one-component fluid with width $0.031 \sigma. $ The dotted curve with empty circles shows the simulation data extracted from ref. \cite{Sciortino}. 
\end{figure}


\begin{thebibliography}{99}
\bibitem{Note} The equation is written  here  for a colloidal mixture, without hydrodynamic interactions  but since the long time limit the final equations for  $f_{ij}(q)$ do not depend on the  diffusion coefficients $D^0_\alpha$ the same behavior holds for atomic systems. Note the missing density factor in $M_{\mu \nu}$ in ref \cite{Fridolin} due to a typing error.
\end{thebibliography}
\end{document}